\documentclass[conference]{IEEEtran}
\ifCLASSINFOpdf
  \usepackage[pdftex]{graphicx}
  \graphicspath{{../pdf/}{../jpeg/}}
   \DeclareGraphicsExtensions{.pdf,.jpeg,.png}
\else
\fi

\usepackage{comment}
\begin{document}
%
\title{City-Identification of Flickr Videos Using Semantic Acoustic Features}

\author{
\IEEEauthorblockN{Benjamin Elizalde, Guan-Lin Chao, Ming Zeng, Ian Lane}
\IEEEauthorblockA{Electrical and Computer Engineering
Carnegie Mellon University Mountain View, CA 94043}
Email: {bmartin1,guanlinc,mzeng}@andrew.cmu.edu, lane@cmu.edu

}


%


\maketitle

\begin{abstract}
City-identification of videos aims to determine the likelihood of a video belonging to a set of cities. In this paper, we present an approach using only audio, thus we do not use any additional modality such as images, user-tags or geo-tags. In this manner, we show to what extent the city-location of videos correlates to their acoustic information. Success in this task suggests improvements can be made to complement the other modalities. In particular, we present a method to compute and use semantic acoustic features to perform city-identification and the features show semantic evidence of the identification. The semantic evidence is given by a taxonomy of urban sounds and expresses the potential presence of these sounds in the city-soundtracks. We used the MediaEval Placing Task set, which contains Flickr videos labeled by city. In addition, we used the UrbanSound8K set containing audio clips labeled by sound-type. Our method improved the state-of-the-art performance and provides a novel semantic approach to this task. 

\end{abstract}


%
\IEEEpeerreviewmaketitle

\section{Introduction}

Multimedia are the fastest-growing type of content on the Internet. Recently, it has become more common to have multimedia tagged by GPS sensors due to the amount of useful information that can be derived from the geo-location. 
A fundamental problem in geo-tagged media occurs when the geo-tags of an image or video are unavailable or unreliable. Therefore, it is essential to estimate the geographical location of the media based on its content. Moreover, the quality of consumer-produced content on the Internet is unstructured, and therefore imposes a challenge for current signal processing and machine learning algorithms, the sheer amount and diversity of the data also promises opportunities for increasing the robustness of approaches on an unprecedented scale. 

The MediaEval\footnote{www.multimediaeval.org} bench-marking initiative recognizes these challenges and motivations and organizes the yearly Placing Task evaluation \cite{choi2014placing}. The bench-mark requires participants to develop algorithms that automatically estimate the geo-location of videos drawn from Flickr\footnote{www.flickr.com} based on textual meta-data, and images and audio extracted from the videos. The accuracies achieved using meta-data have yielded the best results at city-scale followed by the image content and thus, audio content usage has been virtually ignored in all systems. Other reasons are mostly related to the inherent challenges of consumer-produced media. For example, the audio tracks have large variances in length, the content is unstructured, and most of it may not be descriptive of the geo-location. Nevertheless, there are a few papers that show the potential and benefit of utilizing audio. First, Lei presented a conventional audio-based approach for city-identification of videos in \cite{LeiMultimodal}, which utilizes Mel Frequency Cepstral Coefficients (MFCC) features, transformed into GMM-Supervectors and classified using Support Vector Machines (SVMs). The results show performance of 25.3\% Equal Error Rate (EER), which is significantly better than random at 50\% EER. Second, in \cite{Choi2013} Choi presented a human baseline for location estimation for three different combinations of modalities (audio only, audio + video, audio + video + textual meta-data) and compared the results against a machine learning-based system. The study demonstrated cases when humans could effectively identify audio cue for estimating video’s location when the machine algorithm failed. Humans were also effective at inferring the location by combining visual and audio cues when visual cue alone does not carry enough information for the location estimation. Moreover, there is also work on urban soundscapes~\cite{aucouturier2007bag} showing particular acoustic distributions using Bag of Audio Words. On the other hand, these papers ~\cite{elizalde2014audio,barchiesi2015acoustic} aimed to represent a soundtrack as a collection of labeled sounds.

In this paper, we address the task of city-identification of web videos using only the audio content. Our approach consists of computing and using semantic acoustic features to provide a semantic explanation of the identification. The semantic information is given by a taxonomy of urban sounds. For our experiments, we used the MediaEval Placing Task set, which contains Flickr videos labeled by city and the UrbanSound8K dataset containing clips labeled by sound-type. We compared our method to the state-of-the-art performance in terms of EER. 
Our main contributions are: our novel approach achieved by the semantic acoustic features and the improvement of the state-of-the-art performance. 


\section{Corpora}
\label{Corpora}
In this section we describe the two datasets employed in our experiments. 

\subsubsection{MediaEval Placing Task}

The MediaEval Placing Task 2010 dataset contains videos distributed as a training dataset for a multimedia benchmark evaluation. The videos are typical consumer-produced media that contain unstructured data and sometimes the audio-visual content doesn't provide discriminative elements between cities, at least from the human point of view. For example, some videos have been recorded indoors or in private spaces, which makes the Placing Task nearly impossible. The maximum length of Flickr videos is limited to 90 seconds and about 70 \% of the videos are less than 50 seconds.

In the Placing Task, a prediction of the geo-location of a city is successful if its geo-coordinates are within 5 km of the city center. However, in order to compare our results with the state-of-the-art approach presented in \cite{LeiMultimodal}, we replicated their experimental setup. The authors randomly divided the 1,080 videos from the training set into two splits. Train has 541 videos and test has 539 videos. The following 18 cities in Table \ref{cities} were considered because of the predominance of videos belonging to these cities. Note that these videos have no sounds annotations.

\begin{table}
\small
\centering
  \begin{tabular}{ | c | c | c | c | c |}
    \hline
    Bangkok & Barcelona & Beijing & Berlin & Rio  \\ \hline
    Chicago & Houston & London & Rome & Tokyo \\\hline
    Moscow & New York & Paris & Praha &  \\\hline
    Los Angeles & Sydney & San Francisco & Seoul &\\\hline 
  \end{tabular}
\caption{The 18 cities from the Mediaeval Placing Task subset.}
\label{cities}
\vspace{-2em}
\end{table}

\subsubsection{UrbanSound8K}

For our taxonomy of sounds we used the UrbanSound8K dataset \cite{Salamon:UrbanSound:ACMMM:14}. It contains 8,732 audio samples of up to 4 seconds in duration taken from real field recordings. 
The samples include urban sounds from 10 classes: air conditioner, car horn, children playing, dog bark, engine idling, gun shot, jackhammer, siren, drilling, and street music. Since the samples come from field recordings, there are often other sources present in addition to the labeled source. The recordings come divided into 10 stratified subsets, and we used all of the files.

\section{Audio-based City-identification System}
\label{Audio-based City-identification System}

\begin{figure*}
   \centering
     \includegraphics[width=0.9\textwidth]{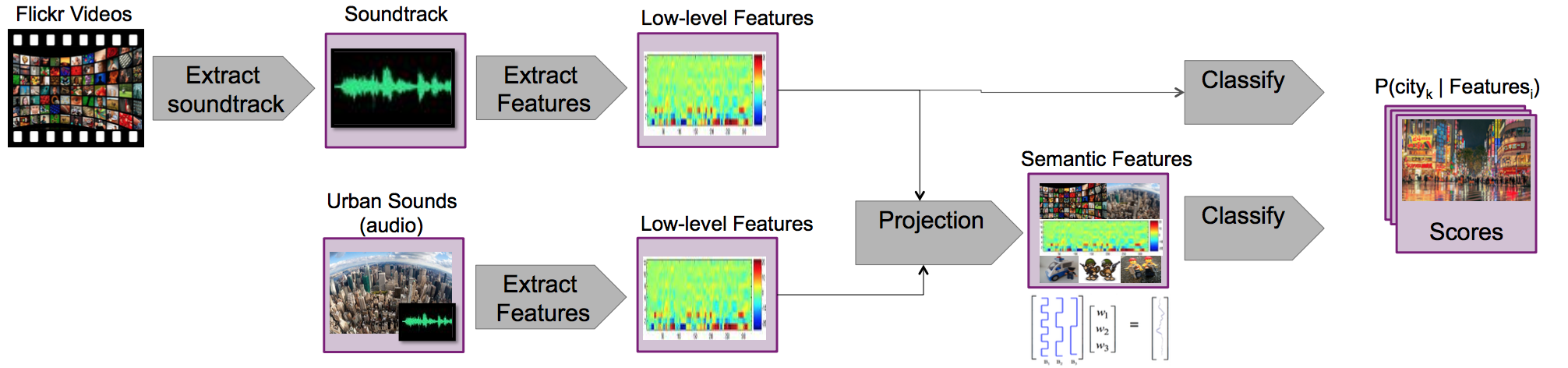}
     \caption{On the top row the system uses low-level features. The soundtrack is extracted from the videos, then low-level features are computed and a classifier was trained and used to identify the videos. On the lower row, the system uses the semantic features. We computed low-level features of the urban sounds and together with the low-level features of the city soundtracks we computed the semantic features. These semantic features were used to train a classifier and perform city identification on the videos.}
     \label{fig:diagramSystem}
     \vspace{-1em}
\end{figure*}
 
Our system is illustrated in Figure \ref{fig:diagramSystem}. On the top row we first extracted the soundtrack from the Flickr videos. Then we extracted low-level features and trained/used a classifier to identify the city of the videos. On the lower row, we computed low-level features of the urban sounds and together with the low-level features of the city soundtracks we computed the semantic features. These semantic features were used to train a classifier and perform city identification on the videos. The details about our low-level features and classifiers are described in Section \ref{Experimental Setup}, however we describe the semantic features in this section due to its importance in our pipeline.

The idea behind our features is to create an approximation of the original signal as a linear combination of bases. 
$$\widehat{Signal} \approx Bases\times Weights = B_1W_1 + B_2W_2 + ... + B_{n}W_{n}$$

We approximate the video's city-soundtracks $(\widehat{Signal})$ with our taxonomy of 10 urban sounds as our bases and their corresponding weights.
$$\widehat{City-soundtrack} \approx carW_1 + sirenW_2 + ... + drillingW_{10}$$

The weights are computed through a projection of the signal onto the bases, and it's achieved with the pseudo-inverse of the bases multiplied by the original signal. The weights matrix has, for each time frame, 10 dimensions each corresponding to the ''presence'' of a urban sound across the soundtrack. The video's city-soundtracks and the urban sounds are acoustic low-level features, which are described in Section \ref{Experimental Setup}. 
$$Weights = pinv(Bases)\times Signal$$

The weights matrix alone is used to provide the acoustic semantic evidence. The matrix could be used as the semantic features for city-identification. However, we obtained better results by instead using the linear combination given by the reconstructed city-soundtrack $(\widehat{Signal})$.

\section{Experimental Setup}
\label{Experimental Setup}

In this section we describe our audio-based city-identification system, including the features and classifiers.

\subsection{Features}
For our experiments we extracted statistical low-level features for the video's soundtrack and for the urban sounds. Additionally, we used two types of semantic features to represent the video's soundtrack. 

\subsubsection{Statistical Features}
We extracted the statistical features based on MFCCs, in a similar way as the one described in \cite{Salamon:UrbanSound:ACMMM:14} since their performance was significantly better than standard MFCCs. First, the MFCCs are extracted on a per-frame basis using a window size of 23.2 ms and 50\% frame overlap. We used 40 Mel bands between 0 and 22050 Hz and kept the first 25 MFCC coefficients (there was no pre-emphasis or liftering applied). Then, the per-frame values for each coefficient were summarized across time using the following summary statistics: minimum, maximum, median, mean, variance, skewness, kurtosis and the mean and variance of the first and second derivatives, resulting in a feature vector of dimension 1 x 275 per audio clip. The problem is that the temporal information of the city soundtracks and the urban sounds is not well represented. Therefore, we extracted the same summary but for every frame using a context information of 45 frames before, the current frame and 45 frames after. Hence, the resulting matrix has as many sliding frames as the MFCCs, instead of only one, and a dimensionality of 275.

\subsubsection{Semantic Features}
The process of extracting the semantic features is described in Section \ref{Audio-based City-identification System}. We used two types of semantic features, the \textit{Weights Matrix} and the reconstruction of the signal given by the \textit{Linear Combination}.

\subsection{Classifiers}
In this paper we used two well-known classifiers, the Random Forest (RF) 
and the Multi-layer Perceptron (MLP).
\subsubsection{Random Forest}
The Random Forest from the \textit{scikit-learn} toolkit~\cite{pedregosa2011scikit} is an ensemble learning approach that constructs a number of decision trees and outputs the average prediction of individual trees. The RF is able to deal well with nonlinear classification tasks. After a tuning stage, we found that 100 trees was the best setting for the Random Forest on this data.

\subsubsection{Multi-layer Perceptron}
The MLP from the \textit{keras} toolkit, is an artificial neural network model. In general, the transformation function (also called activation function) is a non-linear function which projects the input data into a space where it becomes linearly separable. Thus, it addresses well a non-linear classification problem. The architecture of the MLP in our experiments is $input\_dim-1024-1024$. On top of the MLP there is a soft-max classifier for multi-class classification. We used stochastic gradient descent with batch size of 500 examples and a learning rate of $0.0001$. The weight decay momentum parameters are $10^{-6}$ and $0.9$ respectively. The dropout rate is of $0.5$ for each layer. 

\section{Results and Discussion}
\label{Results and Discussion}

\subsection{Features and Classifiers}

\begin{figure}
\includegraphics[width=0.45\textwidth ]{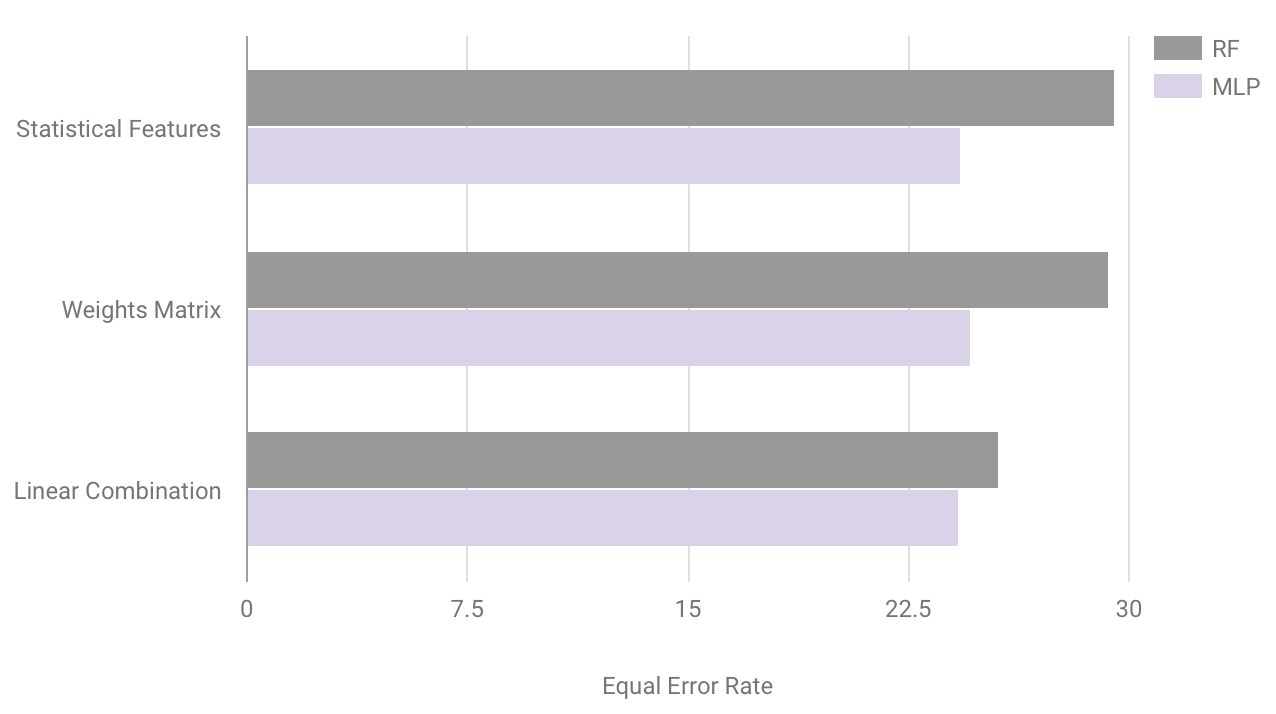}
\caption{The MLP outperformed the RF, which is suggested by a better use of the temporal information in the features. For the low-level features labeled as \textit{Statistical Features}, the EER is 29.5\% for the RF and 24.3\% for the MLP. For the semantic features we have the \textit{Weights Matrix} type with EER 29.3\% and 24.6\%; and the \textit{Linear Combination} type with EER of 25.6\%	and 24.2\%.}
\label{fig:featsvsclassifiers}
\vspace{-1em}
\end{figure}

In Figure \ref{fig:featsvsclassifiers} we present the results obtained from the combination of our features and classifiers. We evaluated our city-identification performance in terms of Equal Error Rate (EER), which is the threshold value where false acceptance rate and false rejection rate are equal. Random guessing produces 50\% EER and a lower number is better.
 
The following results correspond first, to the RF classifier and second, to the MLP classifier. For the low-level features labeled as \textit{Statistical Features}, we achieved 29.5\% and 24.3\% EER. For the semantic features we have two types: the \textit{Weights Matrix} with EER of 29.3\% and 24.6\%; and the \textit{Linear Combination} or approximation of the city-soundtrack with EER of 25.6\% and 24.2\%. These EER results outperformed the 25.3\% presented in the paper \cite{LeiMultimodal} which we are comparing to. In general, the MLP outperformed the RF, which is suggested by a better use of the temporal information in the features as also shown by Ravanelli \cite{ravanelli2015insights} on scenes classification.

The best performance was achieved by the semantic features labeled as \textit{Linear Combination}. This approximation of the city-soundtrack could be seen as a denoised feature, represented by the building blocks of urban sound bases. The other type of semantic features labeled as \textit{Weights Matrix} yielded a higher EER while the loss in performance is not significant. Hence we might as well use the \textit{Weights Matrix} instead and save ourselves the computation associated to the linear combination. 

Our semantic features rely mainly on two qualities. The first is how well the low-level features can represent the bases, because features that capture better the acoustic characteristics of the sounds could improve the projection of the signal into the bases. Therefore, features could improve the semantic evidence and retrieval of sounds. The second is the number and orthogonality of sound-bases, because a larger taxonomy of distinctive sounds provides more bases to better represent the diversity of acoustics within the city-soundtracks as well as extending the semantic evidence. 

\subsection{Semantic Evidence and Retrieval of Sounds}
The semantic features can be used to detect the presence of the urban sounds in the city-soundtracks. In order to show the acoustic content of the 18 cities, we used the \textit{Weights Matrix} features. 

\begin{figure}
\includegraphics[width=0.47\textwidth]{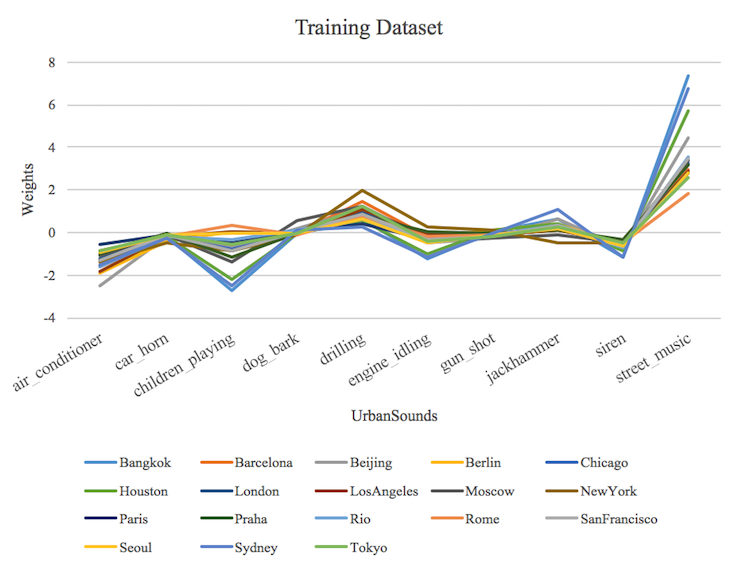}
\caption{The (unnormalized) weights distribution of the 10 urban sounds in the training dataset across the 18 cities. Although the distribution per city might look similar, it is sufficiently different to yield reasonably well city-identification performance. A negative value suggests a null or low presence of the sound.}
\label{fig:distributionTraining}
\vspace{-1em}
\end{figure}

A city level analysis in Figure \ref{fig:distributionTraining} shows the weights distribution of the 10 urban sounds in the training dataset. Although it shows a similar tendency across the cities, 
the distribution per city is sufficiently different to yield reasonably well city-identification performance. 

A urban sound level analysis of the cities required a retrieval approach. Since the MediaEval and UrbanSound8K datasets are independent, there was not any annotation that related both. Therefore, we searched for the presence of sounds in the city-soundtracks based on high weight values. The matrix corresponding to each city-soundtrack was normalized to zero-mean and unit-variance. After manual inspections, we found several successful examples where one or more of the top three urban sounds with highest values were present in the city-soundtrack. For example, in file \textit{12308.mp4} (Barcelona), we found multiple occurrences of \textit{siren} and \textit{car horn}. Moreover, in files \textit{01398.mp4} (San Francisco) and \textit{00429.mp4} (Rio) whose highest weights reflect the presence of \textit{dog bark} and \textit{children playing} respectively. We also found examples with false positives and with near misses. Moreover, we found that \textit{air conditioner} was been triggered when there was background noise and \textit{children playing} together with \textit{street music} were commonly triggered with the presence of speech.




\subsection{Quantity and Orthogonality of the Sounds}

\begin{figure}
\includegraphics[width=9cm,height=4cm]{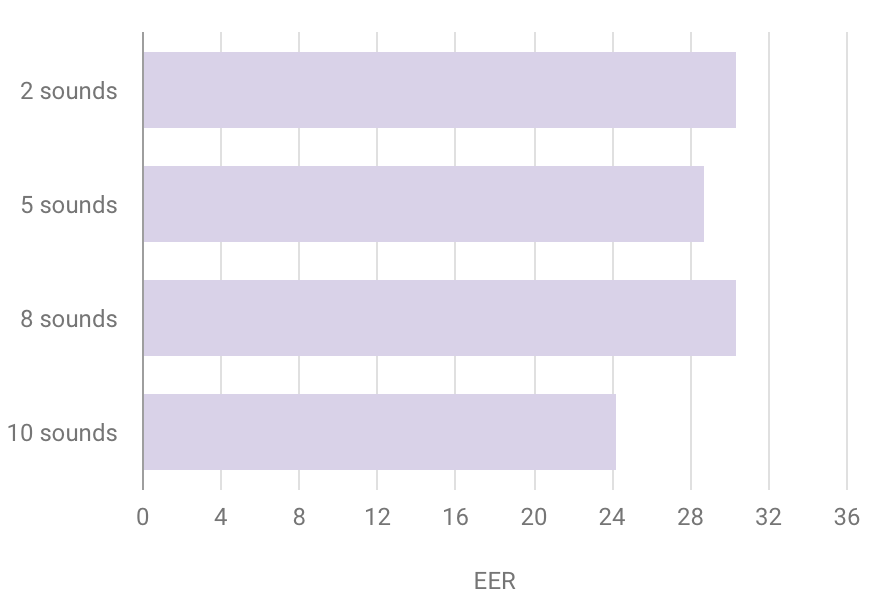}
\caption{The average EER for city-identification improves with the numbers of sound bases. Experiment suggests that more bases could improve city-identification as well as extending the semantic evidence.}
\label{fig:numBases}
\vspace{-1em}
\end{figure}

We wanted to know if the number and distinctiveness of the sound bases would affect the city-identification performance. The intuition is that more distinctive bases would better approximate the variety of acoustics within the city soundtracks. The taxonomy of sounds we used is limited to 10 urban sounds. Therefore, we performed city-identification using our \textit{Linear Combination's} semantic features but with less than 10 sounds bases and the MLP classifier. We chose this setup since it yielded the best results. Since our projection step does not work with a single base due to dimensionality restrictions, we started with groups of two urban sounds and increased the number to five and eight. We randomly picked five groups of two different sounds, then two groups of five different urban sounds and one group of eight different sounds. Each group had unique members. The results can be seen in Figure \ref{fig:numBases} where the average EER for city-identification was 30.4\% for the pairs, then 28.5\% for the quintuples and 30\% for the octuples and 24.2\% for 10 bases as shown in \ref{fig:featsvsclassifiers}. The results show that 10 bases approximates better the city-soundtrack, and thus, a larger taxonomy of sounds could improve the performance. 

\section{Conclusion}
\label{Conclusion}
In this paper, we presented an approach for city-identification using only audio. Our method consists of computing and using semantic features, which offered a performance comparable to low-level features as well as providing semantic evidence of the results. The semantic evidence expresses the relationship between the taxonomy of urban sounds and the city-soundtracks. Moreover, our approach outperformed the state-of-the-art performance and provided a novel take on this task. The analysis of our features suggested potential for sound retrieval, and supported the hypothesis that a larger taxonomy of sounds could improve the identification performance and extend the semantic evidence. For our future work, we plan to increase our taxonomy of sounds and further explore orthogonality of sounds.




\bibliographystyle{IEEEtran}
\bibliography{IEEEfull,sigproc}
%

\end{document}